\documentstyle[prb,aps,epsf,multicol]{revtex}
\newcommand{\bk}  {{\bf k}} % {\bm{k}}
\begin{document}

\draft

%                                              %
%\bibliographystyle{prsty}                     %
%                                              %

\title{Magnetic field effects in energy relaxation
mediated by Kondo impurities}
\author{G.\ G\"oppert$^{1}$, Y.\ M.\ Galperin$^{2}$, 
 B.\ L.\ Altshuler$^{1,3}$, and H.\ Grabert$^{4}$}
\address{
 ${}^{1}$ Physics Department, Princeton University,
        Princeton, NJ 08544, USA \\
 ${}^{2}$ Department of Physics, University of Oslo, 
        P.O.\ Box 1048, N-0316 Oslo, Norway \\
        and Argonne National
        Laboratory, 9700 S. Cass Avenue, Argonne, IL 6043, USA\\
 ${}^{3}$ NEC Research Institute, 4 Independence Way,
        Princeton, NJ 08540, USA \\
 ${}^{4}$ Fakult\"at f\"ur Physik, 
        Albert--Ludwigs--Universit{\"a}t, \\
        Hermann--Herder--Stra{\ss}e~3, 
        D--79104 Freiburg, Germany  
        }

\date{\today}
\maketitle
\widetext

\begin{abstract}
We study the energy distribution function
of quasiparticles in voltage biased mesoscopic
wires in presence of magnetic impurities and
applied magnetic field. The system is described
by a Boltzmann equation where the collision integral
is determined by coupling to spin $1/2$ impurities.
We derive an effective coupling to a dissipative
spin system which is valid well above Kondo 
temperature in equilibrium or for sufficiently
smeared distribution functions in non--equilibrium.
For low magnetic field
an enhancement of energy relaxation is found
whereas for larger magnetic fields the 
energy relaxation decreases again
meeting qualitatively the experimental findings
by Anthore {\it et al.} (cond-mat/0109297). 
This gives a strong 
indication that magnetic impurities are in fact 
responsible for the enhanced 
energy relaxation in copper wires.
The quantitative comparison, however, shows strong
deviations for energy relaxation with small energy 
transfer whereas the large energy transfer regime
is in agreement with our findings.
\end{abstract}

\pacs{73.23.-b, 72.15.Qm, 75.75.+a}
% 73.23.-b Electronic transport in mesoscopic systems
% 72.15.Qm Scattering mechanisms and Kondo effect 
% 75.75.+a Magnetic properties of nanostructures

%                                                      %
\raggedcolumns                                        %
\begin{multicols}{2}                                  %
\narrowtext                                           %
%                                                      %

\section{Motivation and Overview} 
\label{sec:Overview}

\noindent
Energy relaxation of ``hot'' electrons in
disordered conductors at low enough energies
was for a long time believed to be determined by 
direct interaction between electrons. 
Recent experiments on mesoscopic wires
(see Ref.~\onlinecite{PierreERelaxPecs01}
and references therein)
have shown that Kondo impurities 
(localized spins) lead 
to much higher energy relaxation rates
than predicted by the standard 
theory.~\cite{AltshulerEEint85}
% of Altshuler and Aronov

On the other hand the inelastic collisions of
electrons and their spin--flips are directly related
to the phase coherence time probed by weak
localization effects, such as low field 
magnetoresistance.
%A qualitative comparison between these effects 
%based on experiments was reported
%in Ref.~\onlinecite{PierreERelaxPecs01}. 
The problem of decoherence in weak localization
was recently revisited and intensively 
discussed.~\cite{PierreERelaxPecs01,MohantyDecPRL97,MohantyPRL00}

Theoretical studies by various groups
\cite{KrohaASSP00,GlazmanERelaxPRL01,GeorgErelaxPRB01,KrohaPecs01,KrohaNEQPR01,ZawadowskiErelax01} 
lead to a satisfactory and consistent explanation 
of energy relaxation experiments by Kondo impurity
mediated electron--electron interaction
in the gold wires by Pierre {\it et al.} 
\cite{PierreERelaxPecs01}
Those gold wires were contaminated by iron impurities,
and the concentration of the
impurities could be independently estimated from 
the magnetoresistance as well as from the temperature 
dependence of the resistivity. As a result, it was
possible to carry out a parameter free
comparison of theory an 
experiment.~\cite{GeorgErelaxPRB01,KrohaPecs01,KrohaNEQPR01}

At the same time the copper samples in the first energy 
relaxation experiment
by Pothier {\it et al.} \cite{PothierDistrPRL97}
were fitted using the concentration and
Kondo temperature of the paramagnetic impurities 
as free parameters. The parameters obtained from 
energy relaxation disagree with those obtained 
from the 
magnetoresistance experiments of the same 
sample.~\cite{GougamERelaxJLTP00,PierreThesis00}
Therefore, if magnetic impurities 
are responsible for these effects is dubious.

Since the behavior of magnetic impurities is 
sensitive to the applied magnetic field, studies 
of energy relaxation in presence of the 
magnetic field 
could either rule out or validate magnetic 
impurities as relevant scattering process in energy 
relaxation. Recently,
Anthore {\it et al.} \cite{PierreErel01} reported 
results of such experiments in Cu wires that 
indicate a 
strong dependence of the energy relaxation on the
magnetic field suggesting that magnetic impurities 
indeed play a role for the copper wires as well.

In this article we perform a theoretical study 
of transport and energy 
relaxation in a mesoscopic wire in dependence
on an applied magnetic field. We use a 
diffusive Boltzmann equation to account for
the static scatterers and focus in the inelastic 
collision integral on magnetic impurities.
The findings are in qualitative agreement with
the experimental data in
Ref.~\onlinecite{PierreErel01} supporting the
presumption that
scattering with magnetic impurities is
the essential mechanism of energy transfer at
low temperatures. However, the apparent
inconsistency between the values of the
experimentally observed energy relaxation
rate and the dephasing rate extracted from 
the magnetoresistance in Cu wires remains 
puzzling.
%In this article we extend the $t$--matrix approach 
%in Ref.~\onlinecite{GeorgErelaxPRB01} to finite magnetic
%fields using Green's function techniques for 
%non--equilibrium situations covering the old results
%for $B=0$.

Starting with a brief discussion of the 
experiment we
propose in Sec.~\ref{sec:Expsit} a simple physical 
picture to explain the anomalous dependence of the 
energy relaxation on the magnetic field.
In Sec.~\ref{sec:Theordiscr} follows a 
theoretical description
in terms of a renormalized Hamiltonian restricting
the interaction processes to the coupling to 
electron--hole pairs only.
We then present in Sec.~\ref{sec:NumExp} 
the numerical procedure and the comparison with
experimental results. Sec.~\ref{sec:Discussion}
is devoted to a discussion of the 
interpretation and validity of the approach
%where we also point to 
as well as its possible extensions. 
%In Appendix \ref{app:Derivation}
%we derive the collision integral, while the 
%vertex renormalization used in the 
%renormalized Hamiltonian
%is considered in  Appendix \ref{app:VertexRenorm}. 
%Finally, Appendix \ref{app:CorrelationFunction} 
%gives a short derivation of the
%spin correlation functions employed in 
%Sec.~\ref{sec:Theordiscr}.

\section{Experimental situation and 
physical picture}
\label{sec:Expsit}

Here, we briefly describe the experimental situation in
Ref.~\onlinecite{PierreErel01} and a possible qualitative
explanation of their findings.

The experimental setup in Ref.~\onlinecite{PierreErel01} 
consists of a thin copper wire 
of about 45 nm thickness, 105 nm width, and 5 $\mu$m length 
connected with two metallic leads. The leads are biased by 
an external voltage source $U=0.1$ mV and $U=0.3$ mV imposing
a steady state current through the wire.
The setup is placed in a dilution 
refrigerator with a temperature of $25$ mK and a 
magnetic field up to $2.2$ T is applied.
The elastic mean free path can be estimated 
to be much smaller than the length of the 
wire, $L$, so that the transport of electrons 
between the contacts is diffusive. 
%To interpret the results
%it is assumed that longitudinal transport is diffusive, 
%and a proper Boltzmann equation is analyzed. 
The diffusion constant $D$ = 90 cm$^2$/s,
estimated 
from the low temperature resistance, leads
to a diffusion time of $\tau_D=L^2/D=2.8$ ns.
 
The aim of the experiment was to study
energy distribution of electrons. The 
distribution function was determined by tunneling
to an underlying aluminum probe electrode. 
The differential tunneling conductance is 
given by a convolution over the electron distribution 
functions in the wire and in the probe electrode, both 
for $B=0$ and $B\ne 0$, {\it cf.} with  
Refs.~\onlinecite{PothierDistrPRL97,PothierZPB97} and 
~\onlinecite{PierreErel01}, respectively.
In the case of vanishing magnetic field 
the aluminum probe electrode is in the superconducting
state,  and the peaked density of states of the probe 
electrode allows one to straightforwardly extract
the energy distribution in the wire
from the I-V characteristic of the junction.

For finite magnetic fields, however, the aluminum probe 
electrode is in the normal state. 
%and has a finite resistance
In this case the deconvolution  
has been made using a zero bias anomaly. Unfortunately, 
the latter
procedure is less  accurate.
%than in the first case. 
Considering the numerical transformation 
depicted in Fig.~2 of
Ref.~\onlinecite{PierreErel01}, we expect that the strongest 
variations due to uncertainties in the 
shape and the depth of the zero bias anomaly
arise at the ``Fermi points'' 
$\epsilon=\pm eU/2$.
Furthermore, the tunnel probe experiment 
in equilibrium, $U \rightarrow 0$, gave the 
temperature of  $65$ mK
which differs   from the actual refrigerator temperature.
%According to the experimentalists~\cite{PierreErel01},
This might be due to an oversimplification of the 
environmental impedance responsible for the zero bias 
anomaly.~\cite{PierreErel01} 
%This may lead to more uncertainties. 
Consequently, the
resulting distribution function has to be taken with
some care, in particular near the Fermi points.

The electron distribution function in
the absence of inelastic scattering is
just a linear combination of the distributions in the left 
and right electrode~\cite{NagaevPLA92}  
\begin{equation}
 f_0(\epsilon,x)
 = (1-x)f_F(\epsilon-eU/2)+x f_F(\epsilon+eU/2) \, .
\label{eq:initialdistr}
\end{equation}
Here, $f_F(\epsilon)$ is the Fermi function while $x$ is the 
longitudinal coordinate of the observation point in units 
of the total length, $L$. At low temperatures the 
distribution $(\ref{eq:initialdistr})$ has steps in both
$\epsilon$ and $x$ dependencies. These steps are smeared
by inelastic processes, such as 
electron--phonon and electron--electron interaction.  
As a result, the 
distribution in the middle of the wire turns out 
to be almost insensitive to the bath
temperature.~\cite{NagaevPRB95,KozubShotPRB95}
The smearing depends on the effective inelastic relaxation 
time, and the latter can be estimated from the 
experimentally observed distribution function. 

The first experiments~\cite{PothierDistrPRL97,PothierZPB97}  
for $B=0$ have clearly shown that the smearing
is too strong to be attributed to the 
electron--electron or electron--phonon interaction. 
Therefore, in the following we do not
take these interactions into account. 
 
The qualitative outcome of the experiment~\cite{PierreErel01}  
is that the behavior of the inelastic relaxation rate is 
a non--monotonous function of the magnetic field $B$. 
For $B \lesssim B_1=eU/4.3 \mu_B$, the
relaxation rate \emph{increases} with magnetic field 
and reaches a maximum at $B\approx B_1$. 
At stronger fields, $B > B_1$, it decreases with further 
increasing magnetic 
field, and at $B \approx B_2= eU/2 \mu_B$ it reaches almost 
the same value as it had at $B=0$ and then decreases further.

The explanation of such a complex behavior given in 
Ref.~\onlinecite{PierreErel01} is based on electron 
scattering by magnetic impurities. For vanishing 
magnetic field the spin system is degenerate and  
only second or higher order scattering processes
contribute to energy relaxation. For finite magnetic 
fields, there exists also a first order contribution with
energy 
$E_H=g \mu_B B$ (equal to the Zeeman splitting) transferred 
to or from the spin
system. Therefore, the energy relaxation rate increases.
However, for $E_H>eU$ the spins are completely polarized
and can no longer contribute to energy relaxation. 
Consequently, again only higher order processes are 
effective. Comparing this explanation with
the experiment,
one estimates the gyromagnetic factor for the 
impurity spins as $g\approx 2$.

Thus, electron--spin interaction taken into account
in the lowest order of the perturbation theory
explains, in principle, the main experimental features. 
However, from the theoretical point of view, there appears 
a subtlety. The problem is that the higher order terms, 
estimated  within the framework  of the $t$--matrix approach 
in
Refs.~\onlinecite{GlazmanERelaxPRL01,GeorgErelaxPRB01,ZawadowskiErelax01}, 
lead to a divergent contribution 
$\sim J^4/(\epsilon \mp E_H)^2$ to
be integrated over. Here $J$ is the renormalized
coupling constant which defines the strength of the 
electron--impurity
interaction. The
suggestion~\cite{GlazmanERelaxPRL01,KrohaPecs01,KrohaNEQPR01} to
introduce a cutoff at the Korringa width~\cite{KorringaP50}, 
$K\sim J^2$
gives a result that is comparable with the first order 
contribution. Consequently, there is no systematic 
expansion in powers of the coupling constant, and one 
needs a generalized approach which is not based on
expansion in terms of the 
interaction strength.

The aim of the present paper is to develop an approach 
able to treat
lowest and higher order contributions within a unified scheme.
We show that the problem can be described as electrons coupled
to a dissipative spin system. The dissipation of impurity spins
is, in turn,  caused by creation/annihilation of 
electron--hole pairs
due to electron--spin coupling. 
We show that this scheme includes the
divergent higher--order contributions appearing in the
$t$--matrix approach.  
Using our approach, we derive an electron--spin collision 
integral expressed through 
spin--spin correlation functions. An important feature 
of these correlation functions is that their dependences 
on the electron energy is
automatically broadened by Korringa--type processes. These
processes, however, depend on the actual electron distribution 
rather than on the thermal  equilibrium  distribution as in
Ref.~\onlinecite{KorringaP50}. As a result, the applied voltage
plays the role of an effective
temperature.

The crucial difference between the cases $B=0$ and $B \ne 0$ is the
following.
At zero magnetic field all spin--spin
correlation functions are centered at zero energy and behave as 
$K/(\epsilon^2+K^2)$. It is important that in the collision
integral these correlators are multiplied by a combination of the
electron distribution functions which at small
energies is proportional to $\epsilon$. 
Consequently, the broadening turns out to be 
unimportant~\cite{GeorgErelaxPRB01}, and one
can omit $K^2$ in the denominator. This way 
the resulting collision integral becomes
proportional to $J^4$ and one recovers the
results~\cite{GlazmanERelaxPRL01,GeorgErelaxPRB01}  of the
$t$--matrix approach.

At finite magnetic fields, however, the spin--spin 
correlation function decomposes into three contributions. 
The non spin--flip part is still centered at zero energy,
while the two spin--flip contributions are peaked at 
$\epsilon=\pm E_H$.
Let us for a moment accept
the above simplified form of the correlation function and assume
$K \to 0$. Since in this case  
$K/[(\epsilon \mp E_H)^2+K^2]\to \pi \delta(\epsilon \mp E_H)$, 
the two spin--flip correlation functions 
with finite energy transfer lead indeed to the desired first
order in $J^2$ contribution. 
In this simplified case only the non spin--flip correlation 
function contributes to the  
order $J^4$ in the collision integral. 

Using the $t$--matrix approach with a finite cutoff $K$, 
introduced by hand, would lead to a double counting of the 
first order in $J^2$ contribution. Here, on the other hand,
the cutoff is included automatically and the first order
is accounted for correctly.

In this simplified case the reasoning to explain the 
experimental data follows the lines of the experimentalists.
The only difference is that for $E_H>eU$ where the 
spin--flip contribution is already frozen out there remains just
$1/3$ of the $B=0$ energy relaxation because only the non 
spin--flip component contributes to energy relaxation in 
order $J^4$. 
In practice, $K$ is not constant but depends on 
frequency that the correlation functions are not
Lorentzian shaped and therefore the spin--flip terms will 
also contribute to the $J^4$ term. Further, the width 
decreases with 
increasing magnetic field for the non spin--flip 
component and therefore energy relaxation
mediated by magnetic impurities dies out for $E_H \gg eU$. 

We believe that our approach provides a consistent explanation of the
magnetic field dependence of the non--equilibrium electron distribution
in diffusive wires with magnetic impurities.

\section{Theoretical description}
\label{sec:Theordiscr}

Here we present a simplified 
version of the theory with isotropically renormalized coupling 
constant $J$ independent of energy.
In general, the renormalization can be anisotropic 
and energy dependent. These generalizations which do not
alter the underlying physics are discussed in the Appendix. 

We assume that the metallic wire is in the diffusive limit, 
i.e.\ the elastic relaxation time is much smaller than other
time scales. We also assume that the distribution function
of electrons does not depend on the spin. The energy 
distribution of the electrons is governed by the 
Boltzmann equation
\begin{eqnarray}
&&
 \frac{\partial f(\epsilon,x)}{\partial t}
 - \frac{1}{\tau_D} 
   \frac{\partial^2 f(\epsilon,x)}{\partial x^2} 
+I\{f\}=0\, , \label{eq:diff} \\
&&
 I\{f\} 
= \int d\omega 
 \big\{
  f(\epsilon)[1-f(\epsilon-\omega)] W(\omega)
\nonumber \\
&& \qquad \qquad
  -[1- f(\epsilon)]f(\epsilon-\omega) W(-\omega)  \big\}\, .
\label{eq:BoltzmannSimpGen}
\end{eqnarray}
Here, we include the density of states $\rho$ in the 
scattering rate $W(\omega)$ and omit for
convenience the explicit spatial dependence 
of the distribution function. The rate $W$ describes  
the transitions
between two electron states with energies $\epsilon$ and 
$\epsilon-\omega$ mediated by coupling to 
the dissipative spin system. 
Its explicit form is given by
\begin{equation}
 W(\omega)
 =
 (c_{\text{imp}}/\rho \hbar)\, (\rho J/2)^2C(\omega)
\label{eq:RateGen}
\end{equation}
where $c_{\text{imp}}$ is the impurity density.
Further, $C(\omega)$ is the Fourier transform of 
a spin--spin correlation function. The latter can be split as
\begin{equation}  
 C(t)=[C_+ (t)+ C_-(t)]/2 +C_z(t) 
\label{eq:correlfktsplit}
\end{equation} 
where  
\begin{equation}
 C_\pm(t)
 = 
 \langle S^\pm(t)S^\mp(0) \rangle\, , 
\quad 
 C_z(t) 
 =
 \langle S^z(t)S^z(0) \rangle \, . 
\label{eq:correlfkt}
\end{equation}
The averages here mean the spin and electron trace 
weighted with the unknown non--equilibrium density.
The time evolution is 
governed by the Hamiltonian
$H=H_0+H_I$ where 
\begin{equation}
 H_0
 = 
 \sum_{k\sigma} \epsilon_{k\sigma}^{} 
 C_{k\sigma}^\dagger C_{k\sigma}^{} - E_H S^z
\label{eq:freeHam}
\end{equation}
describes free electrons. Here, operators 
$C_{k\sigma}^\dagger$ and $C_{k\sigma}^{}$ 
create and annihilate an electron in a given 
orbital, $k$, and spin, $\sigma$, state. 
$\epsilon_{k\sigma}$ is the energy of this
state. The second term in 
Eq.~$(\ref{eq:freeHam})$ describes a 
spin $1/2$ impurity 
with Zeeman splitting $E_H=g \mu_B B$. The 
interaction Hamiltonian
\begin{equation}
 H_I
 =
 J \sum_{kk'\sigma \sigma'} 
 {\bf S} \cdot {\bf s}_{\sigma' \sigma} 
 C_{k' \sigma'}^\dagger C_{k \sigma}^{}
\label{eq:sdinteracren}
\end{equation}
couples electrons
to the impurity spin system via the renormalized
coupling strength, $J$, rather than the bare one, $J_0$.   
Further, the electron is coupled only to 
one impurity spin since we assume that the impurity 
density $c_{\text{imp}}$ is small enough to neglect 
higher order terms.

Using this renormalized Hamiltonian we will  
restrict our calculation for the time evolution 
of $C(t)$ to coupling to simple 
electron--hole pair excitations only, 
see Appendix \ref{app:Derivation} for details. 
A similar procedure was already used in  
Ref.~\onlinecite{WoelfleZP70} to 
discuss the impurity spin resonance linewidth
in equilibrium using Baym and Kadanoff's kinetic equations.
Here, the spin--spin correlation functions are calculated 
using the conventional
projection operator technique\cite{GrabertPOT82}. 
The results read
\begin{eqnarray}
 C_z(\omega)
& =&
 \frac{1}{2}
 \frac{\nu_z (\omega)}{\omega^2 + \nu_z (\omega)^2}
\label{eq:SzCorrelator} \\
C_\pm(\omega)
 & =&
 \frac{2 {\cal P}_{\pm}
 \nu_\pm  (\omega)}{[\omega \mp E_H]^2 + \nu_\pm (\omega)^2} \,.
\label{eq:SpmCorrelator}
\end{eqnarray}
The functions $\nu_z ,\nu_\pm $ describing damping of spin
fluctuations can be  expressed through an auxiliary function  
\begin{equation}
 \zeta (\omega) 
 =
 \int d\epsilon' f(\epsilon')[1-f(\omega+\epsilon')]
\label{eq:ehpaircoupling}
\end{equation}
stemming from the coupling to electron--hole pairs. They read:
\begin{eqnarray}
 \nu_z (\omega)
 &=&
 \pi (\rho J)^2 
 [{\cal P}_+ \zeta (\omega-E_H)
  +
  {\cal P}_- \zeta (\omega+E_H)]
\label{eq:Korringaz} \\
 \nu_\pm (\omega)
  &=& %\frac{\pi}{4} 
(\pi/4) (\rho J)^2 
 \left[
  2 \zeta (\omega \mp E_H)
  +
  \zeta (\omega) /{\cal P}_{\pm}
 \right] \,.
\label{eq:Korringapm}
\end{eqnarray}
Further,
${\cal P}_\pm$ 
is the
occupation probability for impurity spin up or down, 
respectively. These probabilities are determined by a  
master equation 
\begin{equation}
  \label{eq:me}
  d{\cal P}_\pm/dt
  =
  -\Gamma_\pm {\cal P}_\pm+ \Gamma_\mp {\cal P}_\mp\, , 
  \quad {\cal P}_+ + {\cal P}_-=1    
\end{equation}
which can be solved in the steady state leading to 
\begin{equation}
 {\cal P}_\pm
 =\Gamma_\mp/(\Gamma_+ +\Gamma_-) \,.
\label{eq:OccupProb}
\end{equation}
Here, the inverse lifetime for the spin up/down state, 
$\Gamma_\pm$, is
determined by the expression 
\begin{equation}
 \Gamma_\pm
 =
 \frac{(\rho J)^2}{4 \hbar {\cal P}_\pm} 
 \int d\omega\,  \zeta (-\omega) C_\pm (\omega) \, .
\label{eq:ImpSpinRate}
\end{equation}

For a given electron distribution, the set of 
Eqs.~$(\ref{eq:OccupProb})$ and
$(\ref{eq:ImpSpinRate})$ determines the occupation probabilities 
of the spin system. From Eqs.~$(\ref{eq:correlfktsplit})$ and 
$(\ref{eq:correlfkt})$ one can prove the
sum rule for the correlation function,    
\begin{equation}
 C(t=0)= \int (d\omega/2\pi) \,  C(\omega) = 3/4=S(S+1) \, ,
\label{eq:correlsumrule}
\end{equation}
which is independent of magnetic field. 
In the weak coupling limit 
\begin{equation}
 C_\pm(\omega)
 =
 2\pi {\cal P}_{\pm} \delta(\omega \mp E_H)\, , \quad 
 C_z(\omega)=\pi  \delta(\omega)/2 \, .
\label{eq:correlweakcoupl}
\end{equation}

Inserting this results into the rates 
$(\ref{eq:RateGen})$ and $(\ref{eq:ImpSpinRate})$, 
we recover the Fermi's Golden
Rule expressions for the electron collision operator for the case of
interaction with a single spin. The $\delta$--functions here signalize
energy conservation. The dissipation leads to an energy
uncertainty that, in turn,  results in a broadening of the
$\delta$--functions. The final result differs from that obtained by 
the $t$--matrix approach~\cite{GeorgErelaxBTmatrix01} by allowing 
for a finite energy uncertainty.

\begin{figure}[h]
\begin{center}
\leavevmode
\epsfxsize=0.5 \textwidth
\epsfbox{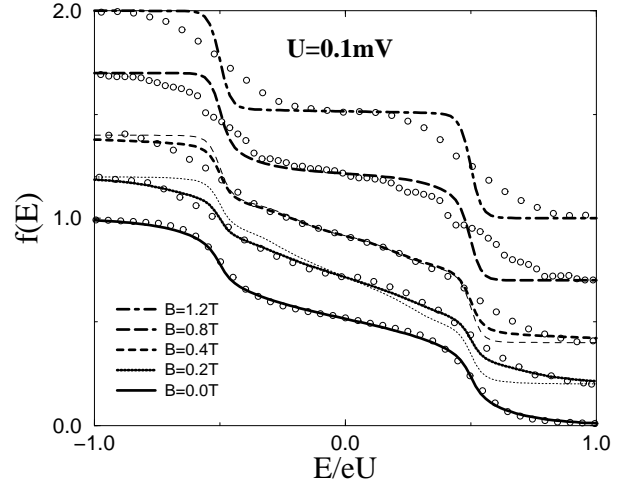}
\caption{Comparison of experimentally determined 
distribution functions from Ref.~\protect\onlinecite{PierreErel01} 
(symbols) with the theoretical
predictions for various magnetic fields and $U=0.1$ mV.
The thick lines are data gained from solving 
the Boltzmann equation where the impurity spins are 
out of equilibrium whereas the thin ones
(differing from the thick lines for two values of B only)
are determined
with impurity spins fixed to equilibrium. 
The distribution functions are given from bottom to 
top for $B= 0.0, 0.2, 0.4, 0.8, 1.2$ T and shifted 
vertically by steps of $0.2$ and $0.3$, respectively.}
\label{fig:fitV0.1}
\end{center}
\end{figure}

For vanishing magnetic field, $E_H=0$, the occupation 
numbers  
${\cal P}_{\pm}=1/2$,
and the correlation function simply reads 
$C(\omega)=3 C_z(\omega)$. Inserted in the collision 
integral, our expression with the renormalized coupling 
constant $J$ from Eq.~(\ref{eq:Jupdwren})
is consistent with the results of
Refs.~\onlinecite{GlazmanERelaxPRL01,GeorgErelaxPRB01}.
The advantage here is that the cutoff,   
$\nu_z (0)=\pi (\rho J)^2 \zeta (0)$, suggested in 
Refs.~\onlinecite{GlazmanERelaxPRL01,KrohaPecs01,KrohaNEQPR01} 
is naturally included. This cutoff equals to 
 the Korringa width $K$.
For weak coupling 
only elastic scattering survives and we get 
$W(\epsilon)=\tau_{\text{sf}}^{-1}  \delta(\epsilon)$ where
the time $\tau_{\text{sf}}$ is usually referred to
as spin--flip time~\cite{HaesendonckPRL87}. 
In this case
the collision integral vanishes identically and one
has to go beyond the lowest order.

Further, 
our expression for the correlation function 
in equilibrium and vanishing magnetic field 
coincides with
the results of W\"oger and Zittartz 
\cite{ZittartzSpindynZP73} based on a Nagaoka--like 
decoupling scheme.
Using Suhl's $t$--matrix for the renormalized coupling constants, 
the spin susceptibilities follow from  proper
kinetic equations.
The procedure is actually similar to the one used in
Ref.~\onlinecite{WoelfleZP70} for 
the analysis of the impurity spin resonance linewidth.

\section{Numerical procedure and 
comparison with Experiment}
\label{sec:NumExp}

For the numerical procedure we use anisotropic
and energy dependent coupling constants 
$J^z_\pm(\epsilon)$ and
$J^\pm(\epsilon)$ given in 
the Appendix \ref{app:VertexRenorm}, 
Eqs.~$(\ref{eq:Jupdwren})$
and $(\ref{eq:Jpmren})$, respectively. 
They have to be calculated self--consistently as 
functionals of the final
non--equilibrium distribution function. 
This procedure 
leads to a slight complication of the formulas in   
the previous section but does not alter their structure. 
The detailed changes to be made are listed in 
the Appendix \ref{app:DetailedTheoDisc}.
Further, in Eqs.~$(\ref{eq:SzCorrelator})$ and 
$(\ref{eq:SpmCorrelator})$ we fix the lifetimes in the 
denominator at resonance, $\nu_z (0)$ and
$\nu_\pm (\pm E_H)$, respectively, and neglect
$\omega$--dependence of $\nu_z$ and $\nu_\pm$. This
approximation formally violates the sum rule 
$(\ref{eq:correlsumrule})$. However, this violation
appears in higher orders in $J$ and thus is not
worrisome. On the other hand in our approximation
the Boltzmann equation $(\ref{eq:diff})$
leads to the effective Fermi distribution in the middle
of a sufficiently long wire, {\it i.e.}, gives the correct 
``hot electron'' limit.

\begin{figure}[h]
\begin{center}
\leavevmode
\epsfxsize=0.5 \textwidth
\epsfbox{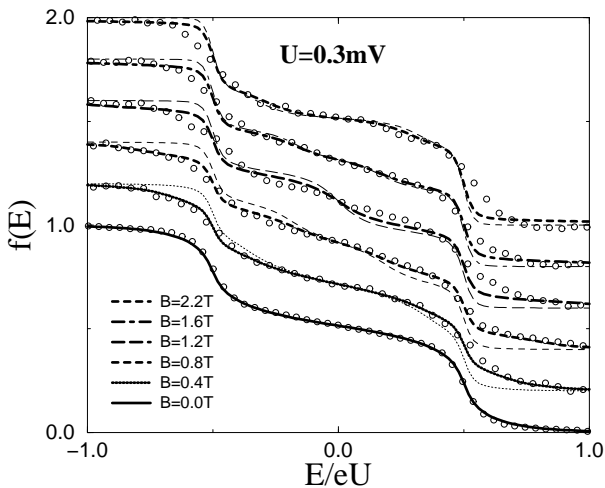}
\caption{Comparison of experimentally determined 
distribution functions from Ref.~\protect\onlinecite{PierreErel01} 
(symbols) with the theoretical
predictions for various magnetic fields and $U=0.3$ mV.
The thick lines are data gained from solving 
the Boltzmann equation where the impurity spins are 
out of equilibrium whereas the thin ones are determined
with impurity spins fixed to equilibrium.
The distribution functions are given from bottom to 
top for $B= 0.0, 0.4, 0.8, 1.2, 1.6, 2.2$ T and shifted 
vertically by steps of $0.2$.} 
\label{fig:fitV0.3}
\end{center}
\end{figure}

We have started with the solution  
(\ref{eq:initialdistr}) of the impurity--free problem 
inserted in Eqs.~(\ref{eq:Jupdwren}) and (\ref{eq:Jpmren}) 
for  the coupling constants, 
Eqs.~(\ref{eq:ehpaircouplingz}) and  
(\ref{eq:ehpaircouplingpm}) for the auxiliary functions 
$\zeta_z (\omega)$ and $\zeta_\pm (\omega)$, and 
Eq.~(\ref{eq:BoltzmannSimpGen}) for  the collision integral. 
Then  
the distribution function was evolved iteratively using the  
Boltzmann equation~(\ref{eq:diff}) with collision
operator (\ref{eq:BoltzmannSimpGen}). 
At each iteration both the coupling constants and the
correlation functions were updated. After about 120 iterations
a stationary solution has been reached.

To make an independent comparison of the finite field data,  
we have fitted  the impurity density with data at $B=0$.
The resulting impurity concentration  $c_{\text{imp}}=8$~ppm is in
accordance with the experimental purity of
copper of $99.999\%$.~\cite{PierreErel01} The
density of states is chosen to be 
$\rho=0.21/$(site$\cdot$eV).~\cite{Kittel96}
The Kondo temperature, not known so far, has been assumed as 
$T_K=0.4$ K.
Further, the gyromagnetic factor 
was chosen as $g \approx 2$, see review of the experimental 
results in Sec.~\ref{sec:Expsit}.

The comparison of the distribution functions for $U=0.1$ mV  
is shown in Fig.~\ref{fig:fitV0.1} and for $U=0.3 $mV 
in Fig.~\ref{fig:fitV0.3}. The symbols
are the distribution functions for several magnetic fields 
obtained from experimental data by a deconvolution procedure.
Note, that a  magnetic 
field of $B=1$ T leads to a Zeeman  splitting 
$E_H \approx 0.12$ meV. Thick lines are
the outcome of our numerical procedure allowing for  the 
non--equilibrium occupation numbers (\ref{eq:OccupProb}) 
for the impurity spins. For comparison,
thin broken lines are the results obtained from
the Boltzmann equation for equilibrium impurity spins.
It is clear that the results for relatively weak magnetic fields, 
$E_H < eU$,
agree with theoretical predictions only if the 
non--equilibrium spin population is taken into account. 
Consequently, we conclude that the
impurity states are indeed out of equilibrium. 
Since the only spin relaxation mechanism taken onto account 
is an interaction with
electrons it follows that ``hot'' impurity centers 
serve as \emph{mediators}
for the electron--electron interaction.

In large magnetic fields, $E_H>eU$, according to 
the theory, the impurity--induced energy relaxation 
is frozen out. Consequently, to obtain a quantitative 
agreement with the experimental results shown in
Fig.~\ref{fig:fitV0.1} one should take into account 
other, though weak, scattering mechanisms.

Further, we observe that the numerical data for 
$U=0.3$mV and $B=1.2$T show an additional step in the 
middle of the energy
distribution function. This can be explained in terms of 
lowest order, $J^2$, scattering with impurity spins where 
the transferred energy between electron and impurity spin
is always $\pm E_H$. The origin is that the Zeeman 
splitting, $E_H$,
at this magnetic field is almost $eU/2$. Electrons in the 
energy region slightly below $-eU/2$ are scattered slightly 
below $\epsilon=0$ because the distribution function differs 
strongly in these two regions. The same statement is correct 
for electrons which scatter from slightly above $\epsilon=0$ 
to slightly above $eU/2$. Scattering between other regions 
is suppressed by the small difference in the distribution 
function at the contributing regions. Using only the first 
order in $J^2$ description for energy relaxation this feature 
would be much more pronounced as the one shown in 
Fig.~\ref{fig:fitV0.3}. The finite width in the correlation
functions $(\ref{eq:SpmCorrelator})$ responsible for
these inelastic processes smear already out most of
the sharp features. We assume that additional energy 
relaxation processes, such as direct electron--electron 
interaction are responsible for the missing step in the
corresponding experimental data. This feature is therefore 
assumed to be an artifact of our numerical calculation caused
by neglecting other scattering mechanisms.

\begin{figure}[h]
\begin{center}
\leavevmode
\epsfxsize=0.5 \textwidth
\epsfbox{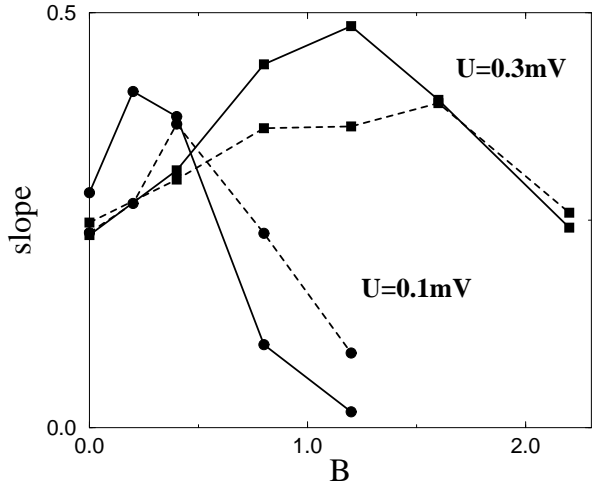}
\caption{Averaged negative slope of the 
energy distribution functions at the plateau near 
$\epsilon \approx 0$ in dependence
of the magnetic field $B$ in units of T. The data
for $U=0.1$mV (dots) are extracted from 
Fig.~\ref{fig:fitV0.1} and for $U=0.3$mV (squares) 
from Fig.~\ref{fig:fitV0.3}. The solid lines
are our numerical data and the dashed lines the experimental
ones.} 
\label{fig:slope}
\end{center}
\end{figure}

Since in our theoretical approach additional scattering 
mechanisms which lead to small energy transfer are 
missing and further the experimental 
data uncertainties are most pronounced at the 
``Fermi points'' $\epsilon = \pm eU/2$ it is not 
reasonable to 
consider these energy regions in more detail. 
To characterize the quality of our fits we focus 
the comparison between our numerical data and the 
experiment to the ``plateau'' region around 
$\epsilon\approx 0$
in the middle of the two--step distribution function. 
The negative slope of the distribution function 
is a good indicator of the energy relaxation strength
as long as it is small. For vanishing 
energy relaxation it is zero and it increases
with increasing energy relaxation.   
In Fig.~\ref{fig:slope} we show the averaged 
negative slope of the energy distribution functions 
at the plateau near 
$\epsilon \approx 0$ in dependence
of the magnetic field $B$. We find 
that the theoretical data (solid lines) meet the 
qualitative outcome in Sec.~\ref{sec:Expsit}
of increasing energy relaxation up to the 
maximum value at about $E_H\approx eU/2$ and then
decreasing again showing at $E_H\approx eU$ about
the same value as at $E_H=0$.
The data, however, lack to explain the experimental 
outcome (dashed lines) in quantitative detail, which 
is not further surprising taking into account the
simplifications we made.

\section{Discussion}
\label{sec:Discussion}

Let us first discuss the outcome of our findings, then 
focus on the justification of various assumptions
not addressed in previous sections, and conclude
with an outlook to further work.

We have succeeded to explain the main features of the 
inelastic relaxation 
due to impurity--mediated electron--electron 
interaction -- effective
relaxation in weak fields and its decrease in 
large magnetic fields, $E_H>eU$.
Further, we found that
the impurity spins are \emph{out of equilibrium}. 
Namely, for $E_H<eU$ both spin states are occupied
even at zero temperature, and the role
of the effective temperature is played by the applied 
voltage. In this sense the impurity centers turn out to 
be ``hot''. Contrarily, when the magnetic field is 
strong, $E_H>eU$, only the 
lowest spin level should be occupied, the occupation of the 
``unfavorite'' spin state is determined by high--order 
processes. Quantitative  
agreement in this region, however, lacks for  small 
energy transfer, responsible for the smearing at the 
``Fermi points''
$\epsilon=\pm eU/2$,
where the experimentally determined distribution functions 
show much  stronger energy relaxation than the one 
expected from the theory. 
It is so far not clear if the small energy transfer 
discrepancy comes from a
failure of our approach in this regime, from other
scattering mechanisms, or is an artifact of the  
deconvolution procedure which was not ruled out
in Ref.~\onlinecite{PierreErel01}.
In the present work we interpret the results using a 
spin $1/2$ model which,
as we believe, is valid at least when the renormalized
coupling constant is small, $\rho J \ll 1$. 
This regime is usually referred to as  
``above the Kondo temperature'' and its meaning
in non--equilibrium situations will be explained in 
the discussion below.

For vanishing magnetic field the experiments 
by Pothier {\it et al.} \cite{PothierDistrPRL97}
and Pierre {\it et al.} \cite{PierreERelaxPecs01}
have shown that the distribution function depends 
only on the ratio $\epsilon/eU$ rather than on 
$\epsilon /k_B T$. This observation is compatible with 
the fact that the Korringa inverse time, $K$, is
proportional to $eU$ at $eU \gg k_BT$. Moreover, using
the results in Sec.~\ref{sec:Theordiscr}
we would even get a distribution function
which depends on the pair of dimensionless
energies, $\epsilon/eU$, and, $E_H/eU$, 
only which qualitatively meets the experimental 
findings. This relation, however, relies on the
assumption that the coupling constant is 
isotropic and energy independent.

\subsection{Inelastic relaxation rate}

Since the distribution functions both of electrons and spins are
out of equilibrium, the inelastic relaxation cannot, in  
general, be discussed  
in terms of a single relaxation rate. In the following we use two
quantities to describe the inelastic relaxation.

%  begin        Georg   new   
First, we consider the collision integral
$(\ref{eq:BoltzmannSimpGen})$ in relaxation
time approximation. We find the corresponding 
rate to be
\begin{eqnarray}
 \frac{1}{\tau_{\text{rt}}}
&\equiv&
 \int d\epsilon' \,  
  W(\epsilon')
  [1-f(\epsilon-\epsilon')+f(\epsilon+\epsilon')]
\nonumber \\
&=&
 \frac{1}{\tau_{\text{sf}}} -
 \int d\epsilon' \,  f(\epsilon-\epsilon')
  [W(\epsilon')-W(-\epsilon')] \, ,
\label{eq:reltimeapprox}
\end{eqnarray} 
with the spin--flip time defined by  
%First, we define an inelastic rate $1/\tau_{\text {in}}$ as a
%contribution to the imaginary part of the self energy due to inelastic
%processes involving magnetic impurities. It can be expressed as,
%cf. with Eq.~$(\ref{eq:BoltzmannSimpGen})$,
%\begin{equation}
% \frac{1}{\tau_{\text {in}}}
% = 
% \int d\epsilon' \,  
%  W(\epsilon')
%  [1-f(\epsilon-\epsilon')+f(\epsilon+\epsilon')] -
%\frac{1}{\tau_{\text{el}}}
% \,.
%\label{eq:inellifetime}
%\end{equation}
%where $1/\tau_{\text{el}}$ is the elastic part of the 
%integral expression. 
%An estimate for the inelastic relaxation rate can be obtained using
the exact relation following from  
the sum rule $(\ref{eq:correlsumrule})$ 
for the correlation function, $C(\omega)$,
\begin{equation}
 \int d\epsilon \, W(\epsilon) 
= 
 \frac{\pi}{2 \hbar} \frac{c_{\text{imp}}}{\rho}  S(S+1)
 \left( \rho J \right)^2  
 \equiv \frac{1}{\tau_{\text{sf}}} \, .
\end{equation}  
%The distribution functions in the integral lead always 
%to a negative contribution which
%leads to a monotonous decrease of the relaxation rate
%with increasing magnetic field. The actual behavior in 
%magnetic field depends also on the occupation numbers 
%${\cal P}_{\pm}$ entering the rates in $W(\epsilon)$.
This relaxation time equals the imaginary part of the 
electron self energy considering only interaction with
localized impurity spins. 
To discuss this expression at finite magnetic 
fields we approximate 
$W(\epsilon)$ by its weak coupling form to get
\begin{equation}
 \frac{1}{\tau_{\text{rt}}}
=
 \frac{1}{\tau_{\text{sf}}}
  \left\{1 - \frac{2}{3}
   \left[ 
    f(\epsilon-E_H)-f(\epsilon+E_H)
   \right]
    ({\cal P_+}-{\cal P_-})
  \right\} 
\label{eq:reltimeapprox1}
\end{equation}
where $({\cal P_+}-{\cal P_-})$ is the mean polarization
of the localized spin, see 
Eqs.~$(\ref{eq:me})$ and $(\ref{eq:OccupProb})$.
For monotonous distribution functions the additional term 
lead to a non--positive contribution and we have always
$1/\tau_{\text {rt}} \le 1/\tau_{\text{sf}}$.
At small magnetic fields the combination of occupation 
numbers and
distribution functions in $(\ref{eq:reltimeapprox1})$ 
decreases and becomes of order $(E_H/eU)^2$. 
In equilibrium we find this contribution to be 
$\sim \tanh^2(\beta E_H/2)$ that leads to an exact 
cancellation of the spin--flip term, 
$1/\tau_{\text {rt}} = 1/3 \tau_{\text{sf}}$,
for large magnetic fields, $E_H \gg k_B T$.
This is in accordance to 
the fact that for large magnetic fields the spin--flip 
contribution is completely frozen out.
Out of equilibrium, the contribution stemming from the
distribution functions behaves similarly and cannot 
explain the increase of energy relaxation for small
magnetic fields.

Whereas in the collision operator $(\ref{eq:BoltzmannSimpGen})$
small energy transfer is canceled by the distribution functions
the relaxation rate in Eq.~$(\ref{eq:reltimeapprox})$  
includes terms stemming from elastic scattering for 
$\epsilon'=0$. To consider the pure inelastic part
we define an inelastic scattering rate by 
\begin{equation}
 \frac{1}{\tau_{\text{in}}}
 \equiv 
 \frac{1}{\tau_{\text{rt}}} -
  \frac{1}{\tau_{\text{el}}}
 =
 \frac{1}{\tau_{\text{rt}}} -
 \int_{-K'}^{K'} d\epsilon' \,  W(\epsilon')
\label{eq:inellifetime}
\end{equation}
with some given cutoff, $K'$, not further specified.

To understand the behavior of the inelastic relaxation time
at small magnetic field one has to recall that the scattering
processes at $B=0$ are almost elastic, a typical energy 
transfer being
smaller or of the order of the Korringa rate, $K$. 
Thus, subtraction of the elastic processes leads to a
drastic decrease of the inelastic relaxation rate, or 
an increase of
the inelastic relaxation time. 
For finite, small magnetic fields, however, the 
inelastic relaxation 
is of order $K+E_H$ and therefore will lead to an 
increase of order
$(E_H/K)^2$. This term has to be compared with the 
monotonous decrease
of order $(E_H/eU)^2$ and dominates in the regime 
$\rho J \ll 1$
explaining the increase of the inelastic relaxation 
rate
for small magnetic fields.

A second way to discuss inelastic relaxation is to 
introduce the
so--called energy relaxation time defined as an average 
energy transfer rate. It is defined as 
\begin{equation} 
  \label{eq:02aa}
 \frac {1}{\tau_\epsilon} 
\equiv 
 - \frac{1}{\bar \epsilon} \, \int
 d\epsilon \, \rho (\epsilon)\,  \epsilon\,  I\{f\}\, . 
\end{equation}
Here $\rho (\epsilon)$ is the electron density of states, 
while
${\bar \epsilon}$ is a constant reference energy to 
normalize the particle energy. A natural scale for 
${\bar \epsilon}$
at $eU \gg k_B T$ is $eU$.
Since $eU \ll D$, where $D$ is the electronic bandwidth, 
the electron density of states can be 
regarded as energy--independent, and after some algebra 
we get
\begin{equation}
  \label{eq:03aa}
 \frac {1}{\tau_\epsilon} 
 = 
 - \frac{\rho}{\bar \epsilon} \, \int
 d\epsilon \int d \omega \,  W(\omega)\, \omega\,
 f(\epsilon)[1-f(\epsilon -\omega)] \, .
\end{equation}
One notices that this expression does not contain elastic
scattering.

\begin{figure}[h]
\begin{center}
\leavevmode
\epsfxsize=0.45 \textwidth
\epsfbox{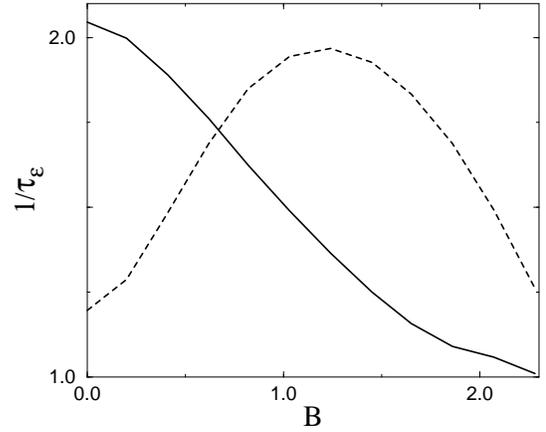}
\caption{Inverse energy relaxation time 
$1/\tau_\epsilon$ (solid line) in arbitrary 
units for $U=0.3$mV depending on magnetic 
field $B$ measured in T. The dashed line 
shows the 
same quantity rescaled by $2/30$ 
where all transferred energies $\omega$ 
in Eq.~$(\ref{eq:03aa})$ 
are added  positively.}
\label{fig:totetrrate}
\end{center}
\end{figure}

In Fig.~\ref{fig:totetrrate} we show the magnetic field 
dependence of the inverse energy relaxation time 
$1/\tau_\epsilon$ in arbitrary units for $U=0.3$mV (solid line). 
We observe a monotonous decrease of the total energy transferred
with increasing magnetic field.
Considering Eq.~$(\ref{eq:03aa})$ as the average  
$(\rho/{\bar \epsilon})\int  d\epsilon d\omega \ldots$ of the
energy transfer, one finds in the integrand,  
depending on the energy,  both positive and
negative contributions.
Although the average $(\ref{eq:03aa})$ monotonously decreases
when $B$ increases,  
the positive and negative contributions in fact increase
for small magnetic fields $E_H \ll eU$.
For comparison we show by the 
dashed line the average $(\ref{eq:03aa})$ where all 
transferred energies $\omega$ 
are added positively. This quantity, in contrast, show
the non--monotonous behavior in dependence of magnetic
field. The physical relevance of this quantity, however, 
is not clear.
This observation  
shows that the notion of a single energy relaxation time
to describe energy relaxation is quite misleading. What happens is
that non--equilibrium spins just \emph{redistribute} the electron
energy between different energy regions, 
the extra energy being transferred
due to diffusion of ``hot'' electrons in real space.

In the experimental papers
Refs.~\onlinecite{PierreERelaxPecs01,PothierDistrPRL97,PierreErel01}, 
as well as in the theoretical ones based upon the  
$t$--matrix approach,
Refs.~\onlinecite{GlazmanERelaxPRL01,GeorgErelaxPRB01,ZawadowskiErelax01},
a contribution of two--particle scattering
processes was also discussed. 
The results are expressed in terms of an effective
interaction strength $\gamma$ which in our notation reads %!!!
$\gamma=(\rho J)^2/\tau_{\text{sf}}$. In the experimental papers 
the results for $B=0$ and
for finite magnetic fields were fitted assuming the same shape of the
electron--electron interaction kernel $\propto 1/\omega^2$. 
According to our findings, the behavior  of the interaction kernel
differs from $\omega^{-2}$.
Consequently, it is difficult to compare their and our
results.

\subsection{Parameters obtained from experiment}

Now let us briefly discuss the parameters obtained from
the fit of the experimental data. Similarly to
Ref.~\onlinecite{GeorgErelaxPRB01} where the 
first experiments\cite{PothierDistrPRL97} with Cu were 
fitted, we
determine the Kondo temperature and the impurity density
(see  Sec.~\ref{sec:NumExp}) from the distribution 
function at $B=0$. These values 
yield
$T_K \approx 0.4$ K and $c_{\text{imp}}=8$~ppm, respectively. 
The procedure proposed in Ref.~\onlinecite{GougamERelaxJLTP00},
based on the analysis of the 
saturated phase breaking time, would
suggest a Kondo temperature below $T \approx 0.1$ K,
see also Ref.~\onlinecite{PierreThesis00}. 
However, assuming such a small $T_K$ in our formalism, 
we would disagree with the experimentally observed  
distribution function at $B=0$ in
Ref. \onlinecite{PothierDistrPRL97}. 
Here, we want to mention that we used the Kondo 
temperature in its simplest form based on the leading 
logarithmic approximation. Corrections may still be
large for this quantity and this disagreement 
could possibly be resolved using a refined version. 

The procedure based on the estimate of the decoherence 
time, $\tau_\phi$, leads to
some other inconsistencies. Namely, an estimation of the 
phase coherence
time with our parameters would lead to 
$\tau_\phi \approx \tau_{\text{sf}}/2 \approx 0.06$~ns 
while, according
to Ref.~\onlinecite{GougamERelaxJLTP00}, 
$\tau_\phi \approx 1$ ns. In return, a fit of the 
decoherence time in Ref.~\onlinecite{PierreThesis00} 
using the same $\tau_{\text{sf}}$ has lead to an
impurity density of $c_{\text{imp}}\approx 0.15$~ppm
much lower than our estimated impurity density.
This means that using the standard theory of weak 
localization \cite{ChakravartyPRep86} the impurity
spins should produce much higher decoherence
rates than observed experimentally. 
The main problem in comparing these two quantities 
may be the following -- the decoherence time is 
measured in equilibrium and it saturates in the case
of Cu below the Kondo temperature if we assume that 
both $\tau_\phi$ and the energy relaxation are determined
by the Kondo impurities. 
In that region, however, our 
treatment, as well as the standard theory of weak 
localization using the spins as random classical objects, 
become inaccurate. 
It is  the aim of future work to develop a theory valid 
below $T_K$. 

In our treatment we used some assumptions which
were not addressed so far.
First, we used the same distribution function for the 
electrons with
different spin projections and 
the standard diffusive form of the Boltzmann
equation. The latter assumption can be justified 
because the elastic
relaxation time\cite{PierreErel01,Kittel96} 
$\tau_{\text{imp}}\approx 0.01$ ps
is much smaller than that for inelastic scattering.
The usage of a single distribution function  
is certainly fulfilled when the spin--orbit relaxation 
rate, $\tau_{\text{so}}^{-1}$,  is larger 
than the rate $\tau_{\text{sf}}^{-1}$ of  
processes which tend to violate the electron spin 
symmetry. This assumption is supported by the results 
of the experiment~\cite{GougamERelaxJLTP00} yielding  
$\tau_{\text{so}} \approx 39$ ps whereas  
$\tau_{\text{sf}}$ is of the order of few
nanoseconds.

We have disregarded some scattering mechanisms to focus
on the magnetic dependence and to discuss the influence
of this sort of impurities only. Direct electron--electron
interaction leads to a pronounced smearing of the distribution
function at low energies. That can be important at rather large   
 magnetic  fields when the energy relaxation mediated by 
magnetic impurities vanishes. Since this additional
effect at $B=0$ is much weaker  than the contribution by  
magnetic impurities it cannot explain the 
deviations from experiment in the region of small  
energy transfer. 

In the Appendix \ref{app:VertexRenorm} we 
determine the renormalized coupling constant 
%$J^z_\pm, J^\pm \cong J$ 
$J$ which replaces the
bare coupling constant $J_0$. The calculations are 
performed within a RPA--like approximation. 
All higher order corrections are supposed to be taken
into account by the renormalization of the
coupling constant $J$. This procedure, as known, 
leads to the so--called overcounting problem which 
has been considered both for the
electron~\cite{SilversteinPR67} and
pseudo Fermion~\cite{ZawadowskiZP69} self energy. 
In these papers an
attempt to cure this problem by setting the energy 
variable
of the renormalized quantity to some special value 
has been made.
No general method that would allow to avoid the 
overcounting has been suggested so far. The only 
message is that as long as
the renormalization is weak and the renormalized 
vertices are changing slowly with energy one can use 
the ``double dressed'' vertices instead
of the ``single dressed'' ones, using the language of 
Ref.~\onlinecite{SilversteinPR67}. Here, we simply adopt
this ``rule'', however, having in mind that if the 
renormalized quantities get too much structure, i.e.\ in 
the Kondo
regime, this approach may become inaccurate.
Our renormalized vertices are similar to those 
obtained using 
the Hamann approximation~\cite{HamannPR67}
in the $t$--matrix approach by
Keiter~\cite{KeiterDispZP76},  which is a generalization
of  Suhl's dispersion approach~\cite{SuhlPR65} to finite 
magnetic fields.
We further neglected the non spin--flip contribution
which may be put into the free Hamiltonian and  
does not influence energy relaxation. 

In the calculation of the correlation functions 
(\ref{eq:SzCorrelator}) and (\ref{eq:SpmCorrelator})
we neglected the Knight shift as well as the frequency and 
Zeeman energy renormalizations.  
The Knight shift just leads to
a simple addition to the ``not really known'' 
Zeeman splitting
of the impurity spin and can be disregarded.
The other terms can be determined by the Kramers--Kronig
relation from the Korringa widths  
(\ref{eq:Korringaz}) and (\ref{eq:Korringapm}). 
They vanish for $\epsilon, E_H =0$
and give just a renormalization of higher order in 
$(\rho J)^2$ which is usually disregarded well  
above the Kondo temperature. However, below the Kondo temperature
these corrections would lead to an anomalous behavior 
and one has to go beyond the lowest order in the 
memory function even if one
uses already renormalized coupling constants, see
Ref.~\onlinecite{ZawadowskiZP69}  for a discussion.

\subsection{Kondo Temperature}

The regime where all our calculations are actually 
valid is defined here as the regime where the lowest 
order in the logarithmic corrections, such as in 
Eqs.~$(\ref{eq:renormfirstpm})$ and 
$(\ref{eq:renormfirstz})$, are still small 
compared to $1$. This is what we actually mean when 
speaking about a ``non--equilibrium'' 
regime or about a situation 
``above the Kondo temperature''. The Kondo
temperature is usually defined by the electron
temperature where perturbation theory starts to fail. 
The terms of the perturbative
expansion explicitly depend on the 
electron distribution function. As a result 
even when the bath temperature is reduced
below $T_K$ the perturbative approach can still
be applied provided that the deviation from 
equilibrium is strong enough. 

In Refs.~\onlinecite{KrohaPecs01}, 
\onlinecite{KrohaNEQPR01},
and \onlinecite{ZawadowskiErelax01} the use of an
effective spatially dependent Kondo temperature 
$T_K^*(x)=T_K^{1/x}/(eU)^{(1-x)/x}$ was proposed for 
electrons energetically 
near the Fermi point $\epsilon=-eU/2$ using the free 
solution $(\ref{eq:initialdistr})$ as a basis of the
renormalization. For electrons
near $\epsilon=eU/2$ one has simply to replace 
$x\rightarrow 1-x$. In these terms the effective 
Kondo temperature in the non--equilibrium situation
is always smaller than the equilibrium Kondo 
temperature.

For an arbitrary distribution function it is not as 
straightforward to calculate the 
renormalized quantities as it is for the free 
solution $(\ref{eq:initialdistr})$. 
Equivalently to the notion of a renormalized Kondo 
temperature we propose to 
use a local temperature $T^*(x)$
describing the non--equilibrium situation. 
The Kondo temperature then 
equals its bulk equilibrium value and the local 
temperature 
is a spatially dependent functional of the distribution 
function. For strong electron--electron interaction or 
equivalently in the middle of a very long wire 
$T^*(x)$ equals the 
analytically determined ``hot electron'' 
temperature~\cite{PothierZPB97,NagaevPRB95,KozubShotPRB95}
independent of the scattering mechanism.
Further, this temperature fulfills the boundary
conditions $T^*(x=0,1)=T_{\rm bath}$. 
Therefore, at the ends of
the wire and for temperatures explored experimentally 
we are below Kondo temperature and our 
approach is inadequate. Here, we assume that the 
boundary regions where our approach fails are narrow 
and do not influence
the distribution function in the middle of the wire.
We believe that this assumption is fulfilled for 
long wires that were studied in the 
experiments.~\cite{PierreERelaxPecs01,PothierDistrPRL97,PierreErel01}
Further, it is obvious that our approach  
becomes more accurate for larger applied voltages 
when the distribution function at $B=0$ is a function 
of $\epsilon/eU$ only. 
This is the reason why we focus more on $U=0.3$ mV.
It turns out (see below for details) that 
for $U=0.1$ mV and smaller voltages our
theoretical approach reaches its limit of validity
within the experimentally explored temperature
range.
 
To be more specific we give an analytic expression 
for the 
effective temperature in the case of weak smearing.
The distribution function then equals the free
solution $(\ref{eq:initialdistr})$ and the smearing
is just included in an effective temperature 
$T_{\rm eff}$ characterizing the energetic width
of the smeared steps. (Here, the 
Boltzmann constant $k_B$ is set to one.) For 
simplicity we restrict 
ourselves to $B=0$ but the generalization is 
straightforward. As a basis of the renormalization 
we use Eqs.~$(\ref{eq:gfunction})$ and 
$(\ref{eq:renormfirstz})$. Following the usual 
poor man's scaling procedure we get 
\begin{equation}
 \rho J(\epsilon)
 =
 \frac{\rho J_0}{1-\rho J_0 g(\epsilon)} 
\label{eq:renormexpl1}
\end{equation}
whereby we adiabatically integrated out all
high energy contributions from $-D$ to 
$\epsilon - \tilde{D}$ and $D$ to 
$\epsilon + \tilde{D}$ letting 
$\tilde{D}$ tend to zero. This procedure at
finite temperature or finite effective temperature
leads to an effective lower energy cutoff which 
is either $\epsilon$, $eU/2$, or $T_{\rm eff}$ and 
follows immediately from
\begin{eqnarray}
 g(\epsilon)
&\approx&
 (1-x) \ln 
 \left[ 
  \frac{D}{|\epsilon - eU/2| + T_{\rm eff}}
 \right]
\nonumber \\
&+&
 x \ln 
 \left[ 
  \frac{D}{|\epsilon + eU/2| + T_{\rm eff}}
 \right]  \, .
\end{eqnarray}
Inserted into $(\ref{eq:renormexpl1})$ we get
for the renormalized coupling constant
\begin{equation}
 \rho J(\epsilon)
\! = \!
 \ln
 \left[
  \frac{(|\epsilon - eU/2| + T_{\rm eff})^{1-x}
        (|\epsilon + eU/2| + T_{\rm eff})^{x} }{T_K}
 \right]^{-1}
\label{eq:renormexpl2}
\end{equation}
with
\begin{equation}
 T_K
 = 
 D e^{-1/\rho J_0} \, .
\end{equation}
For sufficiently weak smearing $T_{\rm eff} \ll eU$ 
mostly electrons near the Fermi points 
$\epsilon \approx \pm eU/2$ contribute. 
We may define a local temperature for electrons depending
on the Fermi point. For electrons at 
$\epsilon \approx eU/2$ we get
$T^*_+(x)=T_{\rm eff}^{1-x} (eU)^x$. The other local
temperature for electrons at $\epsilon \approx -eU/2$ 
reads $T^*_-(x)=T_{\rm eff}^{x} (eU)^{1-x}$.
%\begin{equation}
% T^*_\epsilon(x)
% =
% \left\{
%  \matrix{
%    T_{\rm eff}^{1-x} (eU)^x                 & 
%    \qquad \mbox{for} \qquad                 &
%    \epsilon \approx eU/2                   \cr
%    T_{\rm eff}^{x} (eU)^{1-x}                 & 
%    \qquad \mbox{for} \qquad                 &
%    \epsilon \approx -eU/2                   \cr
%    }
% \right. \, .
%\end{equation}
Here, of course, the effective smearing $T_{\rm eff}$
depends on $x$ and coincides at the ends
of the wire $(x=0,1)$ with the bath temperature 
$T_{\rm bath}$.
Since electrons of both Fermi points contribute to 
inelastic processes one has to average the effective 
temperatures. The simplest average which yields
the correct expression in the middle and the ends
of the wire reads
$T^*(x)=
 (1-x)T_{\rm eff}^{1-x} (eU)^x
 +
 x T_{\rm eff}^{x} (eU)^{1-x}$, because at 
$x=0,1$ only electrons at one Fermi point 
contribute.

In the middle of the wire we get simply 
the requirement $T_{\rm eff} eU \gg T_K^2$ for the
calculations to be valid. If the smearing becomes
stronger and $T_{\rm eff}$ is of the order of $eU$ or 
even larger we get from Eq.~$(\ref{eq:renormexpl2})$ 
the requirement $T^*=T_{\rm eff} \gg T_K$.

On the other hand, we may define an effective Kondo 
temperature by the failure of the perturbation
theory. For electrons near the lower
Fermi point $\epsilon \approx -eU/2$ we write 
\begin{equation}
 \rho J(\epsilon)
 =
 1\Bigg/ x \ln
 \left[ T_{\rm eff}
  \frac{|eU|^{(1-x)/x}}{T_K^{1/x}}
 \right]
\label{eq:renormexplZawa}
\end{equation}
and find the effective Kondo temperature to be 
$T_K^*(x)=T_K^{1/x}/(eU)^{(1-x)/x}$
as proposed in Ref.~\onlinecite{ZawadowskiErelax01}.
The Kondo temperature for electrons near the upper 
Fermi point follows analogously. To our opinion the
introduction of a local temperature is much more
intuitive than a modification of the Kondo 
temperature. It seems to us that the latter 
concept can sometimes be misleading.

\subsection{Conclusion}

In this article we studied energy relaxation mediated by
magnetic impurities above Kondo temperature.
We have shown that the effective 
electron--electron interaction is already included 
in the dissipative nature of the spin system. The finite 
line width in non--equilibrium proportional to the applied 
voltage lead to strong energy relaxation even for vanishing 
magnetic field where the spin states are degenerate.
We succeeded to derive a non--perturbative description
valid for arbitrary magnetic fields where the perturbative
t--matrix approach fails due to divergences in the
two electron scattering rate.

We characterized the efficiency of the energy relaxation 
by the slope of the distribution function at 
$\epsilon \approx 0$. 
The magnetic field dependence of this quantity shown 
in Fig.~\ref{fig:slope} shows a non--monotonous
behavior. For small magnetic field the energy relaxation 
strength is increased and for larger magnetic fields it
decreases again. These features meet qualitatively recent 
experimental data on energy relaxation in mesoscopic
copper wires. Although the detailed comparison of the
energy distribution functions lacks to fit quantitatively
all energy regions, this gives a strong indication 
that magnetic impurities are indeed responsible for 
energy relaxation in copper wires.

Several things remain to be done in order to achieve 
a quantitative description for the energy relaxation 
experiment in the broad regions of temperature and 
magnetic field. To go below the Kondo temperature 
the overcounting problem has to be addressed. 
High--order corrections in impurity density might 
require a more complicated 
kinetic equation including quantum effects. 
Finally, for larger magnetic fields and/or low
concentration of the localized spins other
relaxation mechanisms have to be accounted for.
 
To show that magnetic impurities are indeed present
in copper wires one has to describe all the available
experiments using the same set of parameters.  
Since there are strong deviations to the  
magnetoresistance experiments this is still an
open question.

\section*{Acknowledgments}

The authors would like to thank 
I.~L.~Aleiner, L.~I.~Glazman, F.~Pierre, and H.~Pothier 
for valuable discussions.
Financial support was provided by the Deutsche
Forschungsgemeinschaft (DFG), as well as by Princeton University, 
NEC Research Institute and Argonne National Laboratory.
Further, this research is partly supported  by the US DOE 
Office of Science under contract No.\ W-31-109-ENG-38.

\begin{appendix}

\section{Derivation of the collision integral}
\label{app:Derivation}

We start with the `bare'
Hamiltonian
$H=H_0+H_I$ where the free Hamiltonian is given in
Eq.~(\ref{eq:freeHam}). In terms of pseudo Fermions it reads:  
\begin{equation}
 H_0
 = 
 \sum_{\bk\sigma} \epsilon_{\bk\sigma} 
 C_{\bk\sigma}^\dagger C_{\bk\sigma}^{} 
 + 
 \sum_\beta E_\beta a_{\beta}^\dagger a_{\beta}^{} \,.
\label{eq:freeHamPF}
\end{equation}
Here the creation/annihilation operators  
$C_{k \sigma}^\dagger, C_{k \sigma}^{}$ describe electrons 
while the
operators  $a_{\beta}^\dagger, a_{\beta}^{}$ describe 
pseudo Fermions.
The pseudo Fermion states are specified by $\beta=\pm$ 
having the
energies $E_\pm = \mp g \mu_B B/2$.
The interaction Hamiltonian
$$
 H_I
 =
 J_0 \sum_{\bk \bk'\sigma \sigma'} 
 {\bf S} \cdot {\bf s}_{\sigma' \sigma} 
 C_{\bk' \sigma'}^\dagger C_{\bk \sigma}^{}
$$
is the $s-d$ exchange Hamiltonian with the bare
coupling $J_0$. Expressing spin operators through pseudo Fermions one %!
can rewrite it as
\begin{equation}
 H_I= \sum_{\bk \bk'} J^{\beta' \beta}_{\sigma' \sigma}
 C_{\bk' \sigma'}^\dagger C_{\bk \sigma}^{}
 a_{\beta'}^\dagger a_{\beta}^{}
\label{eq:sdinterac}
\end{equation}
(summation over repeated indices is expected).
In addition, we have to take into account the
operator constraint 
$a_+^\dagger a_+^{}
 +
 a_-^\dagger a_-^{}
 =1$
which can be formally done by a projection 
with the help of a complex chemical potential
\cite{WingreenPRB94}. 

Linear in the impurity density $c_{\text{imp}}$
the angular averaged collision integral for the 
classical Boltzmann equation can be expressed as 
\cite{RammerRMP86,Mahan93}
\begin{eqnarray}
 I\{f\}
 =\frac{i}{\hbar}
  \biggl\{
    f(\epsilon) \Sigma^>(\epsilon)
    +
    [1-f(\epsilon)] \Sigma^<(\epsilon)
  \biggr\}
\label{eq:collisiongen2}
\end{eqnarray}
with  
$\Sigma^{>/<}(\epsilon)=\sum_\sigma \Sigma^{>/<}(\bk \sigma,\epsilon)/2$
where $\epsilon=\epsilon_{\bk\sigma}$ is the spin    
averaged self energy assumed to be independent of the 
angular momentum. $f(\epsilon)$ is the
angular averaged distribution function for electrons
of energy $\epsilon$ where we suppressed the spatial
dependence for convenience. Since the self energy is 
proportional to the impurity density we already replaced
the electron Green's functions by their 
unperturbed form and integrated over frequency to get
the classical form of the Boltzmann equation.
\begin{figure}[h]
\begin{center}
\leavevmode
\epsfxsize=0.35 \textwidth
\epsfbox{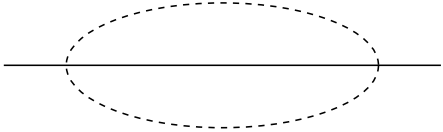}
\caption{Second order self energy graph for the 
electron Green's function}
\label{fig:Sigma2}
\end{center}
\end{figure}
The electron self energy in lowest nonvanishing 
order, depicted in 
Fig.~\ref{fig:Sigma2}, is 
given by a pseudo Fermion bubble
and an electron or hole line in between
\begin{eqnarray}
 \Sigma^>_2(\bk\sigma,\epsilon)
&=&
 c_{\text{imp}}
 J^{\beta' \beta}_{\sigma' \sigma}
 J^{\beta \beta'}_{\sigma \sigma'}
 \sum_{\bk'} \int \frac{d\omega}{2\pi}
             \frac{d \omega'}{2\pi}
   G^>_0(\bk'\sigma',\epsilon-\omega')
\nonumber \\
&&
   {\cal G}^>_0(\beta',\omega+\omega')
   {\cal G}^<_0(\beta, \omega) 
\label{eq:ElSelfEnPT}
\end{eqnarray}
where summation over internal spins is implied.
The self energy $\Sigma^<_2$ is given by the obvious
change of the $<$ and $>$ signs in 
(\ref{eq:ElSelfEnPT}).
Since all corrections in the electron 
Green's functions are of higher orders in $c_{\text{imp}}$
they can always be used in the free form 
$G_0^<(\bk\sigma,\omega)
 =
 2\pi i f(\omega)\delta(\omega -\epsilon_{\bk\sigma})
$ and
$G_0^>(\bk\sigma,\omega)
 =
 -2\pi i [1-f(\omega)]\delta(\omega -\epsilon_{\bk\sigma})
$, respectively.
The pseudo Fermion Green's function in lowest order reads
${\cal G}_0^<(\beta,\omega)
 = 
 2\pi i {\cal P}_\beta \delta(\omega-E_\beta)$
and has to be renormalized in the following. 
Note that
when using the Abrikosov technique~\cite{Abrikosov65}, the occupation
%!!
numbers ${\cal P}_{\pm}$ acquire an 
additional chemical 
potential $\exp(-i\lambda)$ and therefore 
${\cal G}_0^>(\beta,\omega)
 = 
 -2\pi i  \delta(\omega-E_\beta)$, 
see Ref.~\onlinecite{WingreenPRB94}. In contrast to 
Ref.~\onlinecite{WingreenPRB94} we
assume that the occupation probabilities  ${\cal P}_{\pm}$ 
are determined by the non--equilibrium electron distribution. Therefore
they have to be found selfconsistently.
\begin{figure}[h]
\begin{center}
\leavevmode
\epsfxsize=0.4 \textwidth
\epsfbox{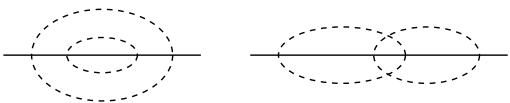}
\caption{Fourth order graphs which are of second order in 
impurity density}
\label{fig:Sigma4_notincl}
\end{center}
\end{figure}
For further proceeding we have to classify the 
appearing graphs. Graphs including two pseudo 
fermion bubbles like shown in Fig.~\ref{fig:Sigma4_notincl} 
are of second order in the impurity density
and are therefore neglected. It is obvious that 
we may include an arbitrary number of additional 
electron--hole bubbles where each one is at 
least of order  $J_0^2$. The two lowest order   
graphs with one 
additional electron--hole bubble are depicted in 
Fig.~\ref{fig:Sigma4}. Considering only these
two topologically different graphs in the 
self energy we obtain the $t$--matrix expression
for the collision integral derived in
Refs.~\onlinecite{GlazmanERelaxPRL01,GeorgErelaxPRB01},
however, with bare coupling constants.
The electron--hole pairs do not only renormalize
the pseudo Fermion propagator, which has been considered in 
Ref.~\onlinecite{ZawadowskiZP69}, 
but also include
bubbles connecting the upper and lower pseudo Fermion
line. The latter graphs are so called crossed diagrams
and lead to vertex corrections~\cite{WoelfleZP70} 
not included in
a simple non--crossing approach.

All other corrections lead to higher orders in $J_0$
for a single electron--hole bubble or outer electron line.
Above the Kondo temperature it is conventionally  accepted that
one may write these corrections 
as a renormalization of the corresponding vertices 
and that in the leading logarithmic order the vertices
are renormalized
independently~\cite{GlazmanERelaxPRL01,ZawadowskiErelax01}.
As a result, two dressed vertices instead of one are used. This
procedure has been explicitly proven  in the leading logarithmic
approximation both for the electron\cite{SilversteinPR67}  and
pseudo Fermion~\cite{ZawadowskiZP69} self energy. An alternative
derivation of the independent vertex renormalization has been given  
in  Ref.~\onlinecite{GeorgErelaxPRB01}.
\begin{figure}[h]
\begin{center}
\leavevmode
\epsfxsize=0.45 \textwidth
\epsfbox{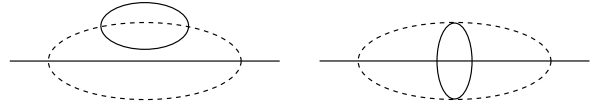}
\caption{Fourth order self energy terms 
including one additional electron--hole bubble.}
\label{fig:Sigma4}
\end{center}
\end{figure}
Assuming this independent renormalization of vertices
to be correct also for more involved graphs and that
this renormalization depends only on the electron energies, 
one can 
equivalently start with a renormalized Hamiltonian
restricting the appearing graphs to simple electron--hole 
bubbles of the order of $J^2$ in the \emph{renormalized} matrix
elements $\tilde{J}^{\beta' \beta}_{\sigma' \sigma}$.

Neglecting the energy dependence of 
$\tilde{J}^{\beta' \beta}_{\sigma' \sigma}$ 
one obtains
\begin{eqnarray}
 \Sigma^>(\bk\sigma,\omega)
\! &=& \!
 c_{\text{imp}}
 \tilde{J}^{\beta' \beta}_{\sigma' \sigma}
 \tilde{J}^{\beta \beta'}_{\sigma \sigma'}
 \sum_{k'} \int \frac{d\omega'}{2\pi}
             \frac{d\hat{\omega}}{2\pi}
   G^>_0(\bk'\sigma',\omega-\omega')
\nonumber \\
&&
 \big\langle
   {\cal G}^>_0(\beta',\hat{\omega}+\omega')
   {\cal G}^<_0(\beta,\hat{\omega})
 \big\rangle_{\rm eh}
\end{eqnarray}
where the average $\langle \ldots \rangle_{\text{eh}}$
means the dressing of the pseudo Fermion bubble with all 
possible electron--hole pairs using renormalized
coupling constants. 
Rephrasing the pseudo Fermions in terms of spin operators
and assuming isotropic coupling $J$
we may write 
\begin{eqnarray}
 \Sigma^>(\omega)
=
 -i \frac{c_{\text{imp}}}{4 \rho} (\rho J)^2
 \int d\omega' C(\omega')[1-f(\omega-\omega')]
\end{eqnarray}
with the time--dependent correlation function $C(t)$ 
given by Eqs.~$(\ref{eq:correlfktsplit})$ and 
$(\ref{eq:correlfkt})$. The self energy $\Sigma^<$
is given by the replacement $f \to 1-f$ and 
$C(\omega) \to C(-\omega)$ which is obvious since
$C(t)$ is an autocorrelation function. Inserting this expression  in 
(\ref{eq:collisiongen2}) we find the collision integral 
in the desired form (\ref{eq:BoltzmannSimpGen}).
The spin--spin correlation functions are calculated then using a
projection operator technique~\cite{GrabertPOT82}. The actual
procedure follows the lines outlined in Ref.~\onlinecite{GrabertAP97}
for the case of two--level systems.

\section{Vertex renormalization}
\label{app:VertexRenorm}

We proceed to determine the renormalized 
coupling constants. They can be derived in different ways:
by a poor man's scaling procedure \cite{AndersonJPC70}, 
Suhl's $t$--matrix approach \cite{SuhlPR65}, 
following Abrikosov and solving a 
generalized vertex equation \cite{Abrikosov65},
or considering lower 
order corrections and assuming similar resummation 
procedure as in equilibrium. According to
Refs.~\onlinecite{SilversteinPR67} and
\onlinecite{DukePR67} 
the renormalized vertex reads
\begin{equation}
 \Gamma(\epsilon' \sigma', \omega' \beta' |
        \epsilon \sigma, \omega \beta ) 
 =
 \Gamma_0 + 
 \Gamma_e(\epsilon+\omega) + 
 \Gamma_h(\epsilon-\omega') 
\label{eq:vertexdef}
\end{equation}
where 
\begin{equation}
 \Gamma_0
 =
 J^{\beta' \beta}_{\sigma' \sigma}
 = 
 J_0 {\bf S}_{\beta' \beta} \cdot {\bf s}_{\sigma' \sigma}
\end{equation}
is the bare vertex of the $s-d$ exchange Hamiltonian.
$\sigma, \beta$ are the incoming electron and pseudo Fermion
spins and $\epsilon, \omega$ are the incoming electron and 
pseudo Fermion energies. The primed quantities are the outgoing 
spins and energies, respectively. 
Energy conservation is fulfilled at each 
vertex meaning $\epsilon'=\epsilon+\omega - \omega'$.
The electron and hole vertex parts can be assumed  
to depend on a single energy variable.~\cite{ZawadowskiZP69} 
The electron vertex $\Gamma_e$
depends on the sum of incoming electron and pseudo Fermion 
energies $\epsilon+\omega$, and the hole vertex part 
$\Gamma_h$  on the difference of the incoming 
electron and outgoing pseudo Fermion energies 
$\epsilon- \omega'$. In the lowest order in 
$J_0$ the retarded quantities read
\begin{eqnarray}
 &&\Gamma^{r}_{e/h}(\epsilon)
 = \frac{i\rho J^{\beta' \gamma}_{\sigma' \kappa }
   J^{\gamma \beta}_{\kappa \sigma}}{2} \!
 \int \! \frac{d\xi}{2\pi} \int \! d\epsilon_{q\kappa}
 \bigg[
   G^{r}_0(q\kappa, \epsilon \mp \xi) 
   {\cal G}^K_0(\gamma, \xi)
\nonumber \\
&& \qquad
   +
   G^K_0(q\kappa, \epsilon \mp \xi) 
   {\cal G}^{r}_0(\gamma, \xi)
 \bigg]
\approx 
 J^{\beta' \gamma}_{\sigma' \kappa }
 J^{\gamma \beta}_{\kappa \sigma}
 \, \rho 
 g(\epsilon \mp E_\gamma) \, . \nonumber
\end{eqnarray}
Here the upper sign stands for the electron vertex and 
the lower sign for the hole vertex, respectively.
The auxiliary function 
\begin{equation}
 g(\epsilon)
 =
 \int_{-D}^{D} d\epsilon'
 \frac{f(\epsilon')-1/2}{\epsilon-\epsilon'+i\delta}.
\label{eq:gfunction}
\end{equation}
leads in equilibrium to the usual logarithmic corrections.

Considering the lowest order vertex corrections
to the self energy in Fig.~\ref{fig:Sigma2} and assuming that 
the pseudo Fermion
energies are fixed to resonance, 
we get 
\begin{equation}
 \Gamma^\pm
 = 
 \frac{J_0}{2}
 \left\{
   1+ \frac{\rho J_0}{2}[
    g(\epsilon \pm E_H) +
    g(\epsilon)  ]
 \right\}
\label{eq:renormfirstpm}
\end{equation}
for the spin--flip component proportional
to $S^\pm$ and
\begin{equation}
 \Gamma^z_\pm
 = 
 (J_0/4)
 \left[
   1+ \rho J_0 g(\epsilon \pm E_H) 
 \right]
\label{eq:renormfirstz}
\end{equation}
for the component proportional to $S^z$ applying
to a spin up/down electron.
This means that the leading renormalization of 
the `up' electron 
is given by the `down' electrons and vice versa.
This anisotropic decomposition of the renormalized
coupling constants has also been found by
Keiter~\cite{KeiterDispZP76}.
Setting the pseudo Fermion energy to resonance 
is  correct only for the lowest order correction in
the electron self energy in Fig.~\ref{fig:Sigma2}.
However, considering corrections to the graphs in 
Fig.~\ref{fig:Sigma4} we find that there the 
pseudo Fermion energies are weighted with a divergent term
$ \propto(\omega-E_\beta)^{-2}$  justifying 
the use of the resonance energy 
$\omega \approx E_\beta$ in the renormalization.
Assuming that this is also justified for more
involved graphs we use these corrections as a 
basis of our renormalization.
%  begin  Georg
To write the renormalized Hamiltonian in terms 
of effective anisotropic and energy dependent 
coupling constants we reintroduce the spin operators
and define 
$J_{\pm}^z(\epsilon)=4 \Gamma_{\pm}^z(\epsilon)$
and 
$J^{\pm}(\epsilon)=2 \Gamma^{\pm}(\epsilon)$.
%  end    Georg
%
% 
\begin{figure}[h]
\begin{center}
\leavevmode
\epsfxsize=0.45 \textwidth
\epsfbox{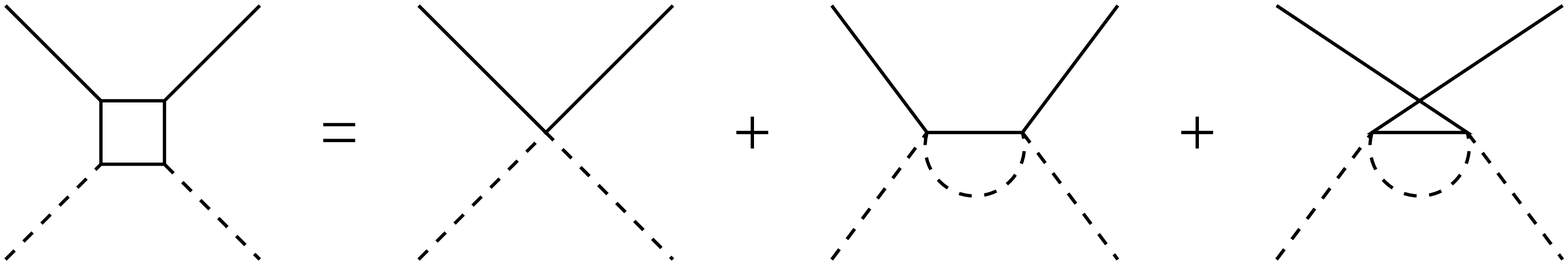}
\caption{Lowest order vertex renormalization}
\label{fig:Vertex}
\end{center}
\end{figure}
To define a `Kondo scale' we 
consider the renormalized coupling constants as 
functions of the incoming electron energy 
and  use the Hamann approximation \cite{HamannPR67} for the 
renormalized quantities. This approximation  yields the same
leading--logarithm expansion as  the poor man's scaling
procedure based on the expressions 
(\ref{eq:renormfirstpm}) and (\ref{eq:renormfirstz}), 
however, taking care of the 
unitarity condition in each order. 
In this approximation the $S^z$ coupling constant reads 
\begin{eqnarray}
 J_{\pm}^z(\epsilon)/J_0 
&=& 
 \Big\{ 
   \left|1
   - (\pi \rho J_0)^2 S(S+1)/4
   - \rho J_0 g(\epsilon \mp E_H)\right|^2 
\nonumber \\
&& \qquad 
   + (\pi \rho J_0)^2 S(S+1)
\Big\}^{-1/2}
\label{eq:Jupdwren}
\end{eqnarray}
in accordance with a high temperature expansion of 
Keiter \cite{KeiterDispZP76}. Analogously
the spin--flip process renormalization reads
\begin{eqnarray}
\label{eq:Jpmren}
 J^\pm(\epsilon)/ J_0 &=&
  \Big\{
    \big| 
     1  - (\pi \rho J_0)^2 S(S+1)/4  - \rho J_0 [g(\epsilon)
\nonumber \\
&& 
\hspace*{-0.7cm}
                                  +  g(\epsilon \pm E_H)]/2 
    \big|^2
   + (\pi \rho J_0)^2 S(S+1)
  \Big\}^{-1/2} \,.
\end{eqnarray}
These quantities coincide for $B=0$ with those
used in Refs.~\onlinecite{GeorgErelaxPRB01,BrenigZP70}
and cover the results obtained by the poor man's scaling 
procedure, {\it cf}.~discussion about Kondo temperature
in Sec.~\ref{sec:Discussion}.
Note that the renormalized coupling constants  
remain finite
also below the Kondo temperature where their meaning 
is questionable.

\section{Calculation of the correlation function}
\label{app:CorrelationFunction}

We briefly present the method used to  
derive the results
in Sec.~\ref{sec:Theordiscr}. Our main interest
is  the spin--spin correlation functions
(\ref{eq:correlfkt}). Since $C(t)$ is an 
autocorrelator 
its Fourier transform is given by the expression 
$C(\omega)=\tilde{C}(-i\omega+\delta)+\tilde{C}(-i\omega-\delta)$
where $\tilde{C}(z)$ the Laplace transform of the
correlator. Using projection operators 
$P^z X=S^z \langle X S^z \rangle/\langle S^z S^z \rangle$ 
for $C_z$ and
$P^\pm X
 =
 S^\pm \langle X S^\mp \rangle/\langle S^\pm S^\mp \rangle
$
for the $C_\pm$ component, one can derive a formally  %!
exact integro--differential equation \cite{GrabertPOT82}
\begin{equation}
 \dot{C}_a(t)
 = 
 \Phi_a C_a(t) 
 -\int_0^t \!\! du\,  \phi_a(t-u) C_a(u)
\end{equation}
with the solution in terms of the Laplace transform
\begin{equation}
 \tilde{C}_a(z)
 =
 \frac{C_a(t=0)}{z-\Phi_a+\tilde{\phi}_a(z)}
\end{equation}
where $a=z,\pm$. Here, 
$\Phi_z=\langle \dot{S}^z S^z \rangle/\langle S^z S^z \rangle=0$
and
$\Phi_\pm
 =
 \langle \dot{S}^\pm S^\mp \rangle/\langle S^\pm S^\mp \rangle
 =
 \mp i \tilde{E}_H
$
leads to the free propagation of the correlators,
where $\tilde{E}_H$ includes the Knight shift neglected 
throughout the article. The averages are calculated 
with the proper 
steady state density defined by the stationary solution
of the Boltzmann equation. 
The memory kernel $\phi_a(t)$ for the correlation function plays a
similar role as the self energy for the Green's function. 
For the $C_\pm$ term we find
\begin{equation}
 \phi_\pm(t)
 =
 \frac{
   \langle \dot{S}^\pm_r(t) \dot{S}^\mp \rangle
      }{
   \langle S^\pm S^\mp \rangle
       }
 +
 \Phi_\pm
  \frac{
   \langle \dot{S}^\pm_r(t) S^\mp \rangle
      }{
   \langle S^\pm S^\mp \rangle
       } \, .
\label{eq:MemoKernelPM}
\end{equation}
Here, the index $r$ in  $S^\pm_r(t)$ indicates 
 that the dynamics of the spin operator
is reduced by the projection. It is determined  by the expression 
$\dot{S}^\pm_r(t)=\exp[i\hat{L}(1-P^\pm)t] \dot{S}^\pm$
with the Liouville operator $\hat L$ acting as $\hat{L}
\hat{X}=[H,\hat{X}]/\hbar$. 
The memory kernel for the $C_z$ correlation function
is simply given by (\ref{eq:MemoKernelPM})
with the replacement $\pm,\mp \rightarrow z$.
Though formally exact, the above equations  do not allow 
a simple calculation of the correlators. Here, we 
expand the kernel up to  second order in the
renormalized coupling $J$. Since the dynamics of
the expanded kernel function is oscillatory 
the Fourier transformed correlation function
has always the simple form 
\begin{equation}
 C_a(\omega)
 =
 \frac{2 C_a(t=0) \, {\rm Re} \, \phi_a(\omega)}
 {[\omega -i\Phi_a + {\rm Im}\, \phi_a(\omega)]^2 
  +[{\rm Re} \, \phi_a(\omega)]^2} 
\label{eq:CorrelGen}
\end{equation}
with $a=z,\pm$. Similar to  
Green's functions, this 
reflects the fact that in the steady 
state the retarded and advanced 
self energies are complex conjugate,
leading to a similar structure of the 
``$>$'' or ``$<$'' Green's functions. Note that the
roles of the imaginary and real parts of  the memory kernel are
opposite to those of the Green's functions because
of different set of definitions.
Further, we define
${\rm Re}  \, \phi_a(\omega)
 \equiv
 {\rm Re}\, \{\tilde{\phi}_a(-i\omega+\delta)\}$ and the
imaginary part ${\rm Im} \, \phi_a(\omega)$ follows from a 
Kramers--Kronig relation. 
With $C_z(t=0)=\langle S^z S^z \rangle =1/4$ and 
$C_\pm(t=0)
 =\langle S^\pm S^\mp \rangle 
 ={\cal P}_{\pm}
$
and further neglecting the imaginary parts in the 
denominators which lead
to a frequency and Zeeman energy renormalization, 
we find
the expressions $(\ref{eq:SzCorrelator})$ and 
$(\ref{eq:SpmCorrelator})$ for the correlation
functions.

\section{Detailed theoretical description}
\label{app:DetailedTheoDisc}

Starting with 
anisotropic and energy dependent coupling constants
derived in Appendix \ref{app:VertexRenorm} the 
interaction Hamiltonian in terms of impurity spin
operators reads
\begin{eqnarray}
 H_I
&=&
 \frac{1}{2} \sum_{kk'}
 \bigg\{
   S^+ J^+(\epsilon_{k \uparrow}) 
   C_{k' \downarrow}^\dagger C_{k \uparrow}^{} +
   S^- J^-(\epsilon_{k \downarrow}) 
   C_{k' \uparrow}^\dagger C_{k \downarrow}^{}
\nonumber \\
&&+ 
   S^z
   \left[
     J^z_+(\epsilon_{k \uparrow}) 
     C_{k' \uparrow}^\dagger C_{k \uparrow}^{}  -
     J^z_-(\epsilon_{k \downarrow}) 
     C_{k' \downarrow}^\dagger C_{k \downarrow}^{}
   \right]
 \bigg\} \,.
\label{eq:sdinteracren1}
\end{eqnarray}
Considering the electron self energy in the 
Boltzmann equation the scattering rate $W$ 
in the collision integral 
$(\ref{eq:BoltzmannSimpGen})$ depends on both, 
incoming electron energy $\epsilon$ and 
transferred energy $\omega$
\begin{eqnarray}
 W(\epsilon, \omega)
&=&
 (c_{\text{imp}}\rho/8 \hbar)\, 
 \big\{
   J^-(\epsilon)J^+(\epsilon')C_+(\omega)
\nonumber  \\
&& \hspace*{1.65cm} +
   J^+(\epsilon)J^-(\epsilon')C_-(\omega) 
\label{eq:RateGen1} \\
&& \hspace*{-0.3cm}  +
   \left[
      J^z_+(\epsilon)J^z_+(\epsilon') 
     +J^z_-(\epsilon)J^z_-(\epsilon')
   \right] C_z(\omega)
 \big\}
\nonumber
\end{eqnarray}
with the outgoing electron energy 
$\epsilon'=\epsilon-\omega$, {\it cf.} 
Appendix \ref{app:Derivation}
and Eqs.~$(\ref{eq:correlfktsplit})$ and
$(\ref{eq:correlfkt})$.
The calculation of the correlation 
functions follows the lines in 
Appendix \ref{app:CorrelationFunction}
and the general form 
$(\ref{eq:CorrelGen})$ leading the 
Eqs.~$(\ref{eq:SzCorrelator})$ and 
$(\ref{eq:SpmCorrelator})$ remain the same.
The changes are only in the damping 
\begin{eqnarray}
 \nu_z (\omega)
 &=&
 \pi \rho^2 
 [{\cal P}_+ \zeta_+ (\omega-E_H)
  +
  {\cal P}_- \zeta_- (\omega+E_H)]
\label{eq:Korringaz1} \\
 \nu_\pm (\omega)
  &=& %\frac{\pi}{4} 
(\pi/4) \rho^2 
 \left[
  \zeta_z (\omega \mp E_H)
  +
  \zeta_{\mp} (\omega) /{\cal P}_{\pm}
 \right] 
\label{eq:Korringapm1}
\end{eqnarray}
with the auxiliary functions
\begin{eqnarray}
 \zeta_z (\omega) 
\! &=& \!\!
 \int \!\!  d\epsilon' 
 \left[
   J^z_+(\epsilon')J^z_+(\omega+\epsilon') 
  +J^z_-(\epsilon')J^z_-(\omega+\epsilon')
 \right]
\nonumber \\
&& \hspace*{1cm}   \times
 f(\epsilon')[1-f(\omega+\epsilon')] 
\label{eq:ehpaircouplingz}    \\
 \zeta_\pm (\omega) 
\! &=& \!\! 
 \int \!\! d\epsilon' 
   J^\mp(\epsilon')J^\pm(\omega+\epsilon') 
 f(\epsilon')[1-f(\omega+\epsilon')] \, .
\label{eq:ehpaircouplingpm}  
\end{eqnarray}
The corresponding master equation 
$(\ref{eq:me})$ for
the occupation probabilities ${\cal P_\pm}$
remains the same but the rate changes to
\begin{equation}
 \Gamma_\pm
 =
 \frac{\rho^2}{4 \hbar {\cal P}_\pm} 
 \int d\omega\,  \zeta_\pm (-\omega) C_\pm (\omega) \, .
\label{eq:ImpSpinRate1}
\end{equation}
We used these formulae for the numerical 
determination of the distribution function 
in Sec.~\ref{sec:NumExp}, however, the 
main features do not alter even if one uses the 
simplified description in 
Sec.~\ref{sec:Theordiscr}.

\end{appendix}

%                                         %
%\bibliography{lit}                       %
%                                         %

%                                         %
\end{multicols}                          %
%                                         %

\end{document}